\documentstyle[aps,preprint,tighten,epsf]{revtex}
\begin{document}
\preprint{\begin{minipage}{3.5cm}HUB-EP-98/58\\
GUTPA-98-09-01\\\end{minipage}}
\draft
\title
{A lattice potential investigation of quark mass and volume dependence
of the $\Upsilon$ spectrum}
\author{Gunnar S.\ Bali\thanks{Electronic address: bali@physik.hu-berlin.de}}
\address{Institut f\"ur Physik, Humboldt-Universit\"at zu Berlin,
Invalidenstra\ss{}e
110, 10115 Berlin, Germany}
\author{Peter Boyle\thanks{Electronic address: pboyle@physics.gla.ac.uk}}
\address{Department of Physics and Astronomy, The University of
Glasgow, Glasgow
G12 8QQ, Scotland}
\date{\today}
\maketitle

\narrowtext
\begin{abstract}
We investigate bottomonia splittings by solving a
Schr\"odinger-Pauli-type equation with 
parametrisations of QCD potentials around those that have been
determined previously in lattice simulations. This is done both,
in the continuum and on finite lattices with resolutions ranging
from $a=0.2$~fm down to $a=0.025$~fm and extent of up to 12 fm or
$144^3$ lattice points. We find a strong dependence of some
splittings, in particular the $2S-1S$
and $1P-1S$ splittings, on both the quark mass and
the short range form of the static potential in the neighbourhood
of the $b$ quark mass, while splittings such as $3S-2S$
and $2P-2S$ show reduced dependence on the short distance potential.
We conclude that the quenched quarkonium spectrum cannot be
matched to experiment with a simple redefinition of the lattice spacing.
We investigate the size of relativistic corrections as a function
of the quark mass.
Finite size effects are shown to die out rather rapidly as the
volume is increased, and we demonstrate the restoration 
of rotational symmetry as the continuum limit is taken.

\end{abstract}

\section{Introduction}

Potential models
have been applied to explain quarkonium spectroscopy~\cite{appel}
in cave paintings dating from the mesolithic era of 
QCD~\cite{potmod1,potmod2,potmod3,potmod4,potmod6}.
In the beginning, they were based on parametrisations of the
potential~\cite{potmod1,potmod2,potmod5}
which were --- at best --- QCD motivated
and subsequently fitted to 
the observed spectrum. During the past few years, however,
a two-body Schr\"odinger-Pauli-type Hamiltonian incorporating
the static potential and
relativistic corrections due to spin dependent interactions as well
as perturbations of the heavy quark propagator around the 
static Wilson line has been derived 
directly from the QCD Lagrangian to order $v^4$ in the heavy quark
velocity~\cite{EF,EFG,BBMP,baliwachter}. 
One-loop matching coefficients with renormalisation group running
between terms within the effective
Hamiltonian and QCD have been
obtained~\cite{chen,baliwachter,balzereit,manohar,soto}.
Non-QCD input has still been required until all
non-perturbative potentials have recently been determined
from quenched lattice QCD by one of the
present authors~\cite{baliwachter,baliwachter1}.
Parametrisations have been fitted to these lattice potentials
enabling Schr\"odinger-Pauli-type solutions to be formed
that are genuine predictions of (quenched) QCD.
The spectroscopy for quarkonia shows similar behaviour to that obtained
using quenched NRQCD; however due to the use of a Hamiltonian
formulation of the problem
the potential approach allows the calculation of a wider range of
states than it is possible within NRQCD,
in a manner that is both accurate and more
robust against finite size effects
(FSE).
Moreover, once the Hamiltonian is obtained, the quark mass can easily be
varied.

We would like to mention that so far retardation effects have been
neglected within the potential approach\cite{volleut}.
However, it appears
likely that such effects, which
have their origin in radiation of ultra-soft gluons, can be
incorporated into the potential formulation in a rather elegant
and straightforward
manner~\cite{pNRQCD}
that should automatically account for spin exotic
states and mixing between quark model and hybrid states~\cite{bali}.
This work indicates that --- at least for the lowest lying
bottomonium states in the quenched approximation
--- such effects are tiny, which explains the good
agreement of our results with Lattice NRQCD predictions. 

The obvious drawbacks of the potential approach are
(a) that it cannot easily be extended to higher orders in
the velocity where more and more non-potential type contributions will
have to be included and (b) that --- unlike HQET/NRQCD ---
it only applies to heavy-heavy systems.
Bottomonia systems are the ideal application and ---
having access to all excited states --- we are indeed provided
with much more insight than if we were just able to calculate the
lowest few excitations.

In this paper we shall investigate the major sources of error in lattice
calculations of the bottomonium spectrum.
For simplicity, we will only employ Cornell-type parametrisations
of the potentials, neglecting a weakening
of the effective QCD coupling at small quark separation. 
We study finite volume and discretisation effects by solving the order $v^2$
Hamiltonian on (three-dimensional) lattices, varying the physical
volume and lattice spacing.
Moreover, we study the size of relativistic and radiative
corrections and estimate possible quenching effects by simulating
the order $v^2$ and order $v^4$ Hamiltonians in the continuum,
using parametrisations of the potentials that are motivated by fits to
quenched lattice data~\cite{baliwachter}.
We shall argue that, in order to reproduce the observed spectrum correctly,
the unquenched potential must
have a significantly stronger effective Coulomb coefficient at
intermediate distances than
determined within the quenched approximation, even when ignoring
the weakening of this coefficient with the distance in the latter case.

The principal effect of this is that observable quantities
which exhibit a large sensitivity to the short distance
potential (e.g. states of small physical extent or fine splittings)
will yield results that are inconsistent with those obtained
from states that are more sensitive to physics at larger distances.
These discrepancies will depend on the mean radius of the state in a 
manner that is far less benign than the na\"\i{}ve
perturbative expectation of an enhanced
slope of the logarithmic running of the coupling
in the quenched approximation would suggest.

The paper is organised as follows.
In Sec.~\ref{Expect} we present some simple theoretical expectations
for the spin averaged spectrum and comment on results from Lattice
spectroscopy investigations in view of these general considerations.
In Sec.~\ref{Method} we discuss the methods used for solving
the Schr\"odinger-Pauli equation both in the continuum and
using a discrete lattice formulation with boundary
conditions imposing various cubic group representations to fix the
orbital angular momentum, before we present our results in Sec.~\ref{Result}.
We start with a discussion of systematic uncertainties imposed by
truncating the expansion of the QCD Lagrangian at a given order in $v$ as well
as due to radiative corrections to the matching coefficients of
operators of dimension higher than four. Subsequently,
results on the quarkonium spectrum as a function of the
inverse quark mass with various effective Coulomb coefficients
around those suggested by quenched and partially unquenched Lattice
simulations are presented. The implications
on the expected $n_f = 3$, $m_u \approx m_d \approx 5$~MeV,
$m_s \approx 100$~MeV
Coulomb coefficient are discussed, and we rationalise both, the calculated
mass dependence of the quenched $1P\,1S$ splitting and the analysis methods
common in lattice heavy quark simulations.
Finally, we proceed to a discussion of the dependence of the results
on the lattice volume and spacing and characterise the likely finite 
volume errors and rotational symmetry
restoration properties in the continuum limit of the lattice theory.
In Sec.~\ref{Summary}, we summarise our results and present a brief outlook.

\section{General expectations}
\label{Expect}
Na\"\i{}vely one might expect that prior to a lattice simulation the
lattice resolution $a$, at which the real world will be studied,
is selected. In practice, however, the theory is non-dimensionalised and
simulated at a given value of the bare coupling.
Subsequently, the dimensionful lattice
spacing is determined by matching an observable to its experimental value.
If this observable depends on quark masses, the process of fixing
the lattice scale cannot be disentangled from the problem of
adjusting the quark masses, and simultaneous matching of two or more
quantities to experiment has to be performed.
It is therefore common to either choose a non-hadronic quantity,
such as~\cite{sommer} $r_0$ (which is related to the
static QCD potential) or to choose quantities defined at almost vanishing 
quark mass, such as $f_\pi$, to alleviate this problem. Despite
the significant difference between the mass of a charm and a bottom
quark, the spin averaged $1P\,1S$ splittings of 
the $\Upsilon$ and $J/\psi$ systems are found to be
essentially the same in experiment. Therefore,
it has been suggested that, in the above mentioned sense, this
splitting was a particularly
good quantity to set the lattice scale $a$~\cite{aida}.

We note that in the potential picture the 
spin averaged splittings scale as $\Delta m \propto m^{-\frac{1}{3}}$ for
a purely linear potential and as $\Delta m \propto m $ in the pure Coulomb
case. For a logarithmic potential, the splittings are independent of
the quark mass (see e.g.\ Ref.~\cite{potmod4}).
With a string tension $\kappa\approx 450$~MeV transition from one
of these limits to the other sets in at a separation of order 0.5~fm,
the static potential in the region $0.2$~fm$<r<1$~fm being approximately
compatible with a logarithmic parametrisation, which provides an
explanation of the similarity
between the spin averaged
charmonia and bottomonia level splittings.

It has been reported in quenched simulations, however, that the $1P\,1S$ 
splitting shows a significant slope as a function of~\cite{boyle,aoki}
$m_\Upsilon^{-1}$. Moreover, both
the $2S\,1S$ and $1P\,1S$ splittings are found to be
almost $20\%$ smaller than the experimental values
if one sets the lattice spacing from light hadronic
quantities~\cite{aida,nrqcd2,manke,ctd,trottier}
like $f_\pi$ or $m_\rho$ or the square root of the
string tension 420~MeV$<\sqrt{\kappa}<$~450~MeV,
as determined from Regg\'e trajectories.
Furthermore, calculations over a mass range extending above bottom
have displayed a steep rise in the splittings~\cite{aoki} in 
the $m \rightarrow \infty$ limit, starting somewhat above the $b$ quark.
{}From the potential picture, this can be explained as follows:
as one increases the quark mass beyond the bottom mass, the states
under consideration move deeper into the Coulomb region and
splittings will diverge linearly in $m$, while in the limit of
light quark masses splittings should grow like $m^{-1/3}$. This
suggests that the beauty and charm masses lie within a broad
minimum of the $1P\,1S$ and $2S\,1S$ splittings, the extent of
experimental agreement largely being accidental. Of course in the
limiting cases themselves, the potential picture will not be applicable: at
light quark masses the description as a non-relativistic system
breaks down while at extremely heavy quark masses, retardation
effects will eventually dominate the dynamics~\cite{volleut}.

It has been argued that
the relevant physical momentum scales that affect bottomonia masses 
differ from those of light hadronic quantities.
Neglecting 
sea quark loops in the quenched approximation alters
the running of the QCD coupling, thereby resulting in
large quenching errors
when comparing quantities determined by different momentum
scales~\cite{aida,ctd,trottier,sesam}.
Indeed, in the presence of two light flavours of
dynamical fermions the disagreement between
mass ratios determined on the lattice and in experiment
appears to be reduced~\cite{sesam,ctd}.
Some authors, therefore, suggest
to adapt the lattice scale $a$ to the system 
under consideration in order to reduce quenching
uncertainities~\cite{aida}.

The ``different lattice spacings for different systems'' argument
obviously reduces the predictive power of quenched and
partially unquenched simulations while
it does not even work
consistently within the $\Upsilon$ system:
by integrating a Schr\"odinger-Pauli
equation with quenched
lattice potentials, the $\Upsilon$ $2S\,1S$
and $1P\,1S$ have been found~\cite{baliwachter}
to be significantly smaller than one would have expected from
the value of $am_{\rho}$. Deviations between experimental
and quenched lattice ratios between $m_{\rho}$ and
splittings incorporating
larger radial states like $3S\,2S$ or $2P\,1P$,
however, are found to be small~\cite{baliwachter,sesaminprep}.
The first observation is in agreement with Lattice NRQCD
results~\cite{aida,sesam}.
Unfortunately, NRQCD precision results
are only available for the $1S$, $1P$ and, to a lesser extent, $2S$ states,
such that the above mentioned inconsistencies within the quenched
$\Upsilon$ spectrum are largely hidden behind
statistical errors and, possibly, FSE for
$3S$ or $2P$ states.

We remark that the relevant
exchange momentum will in general not only depend on
the system but also on the state under consideration.
One should also keep in mind that
within the non-perturbative regime universality of the QCD
$\beta$-function is lost.
Therefore, interpreting differences between mass ratios determined within
the quenched approximation and in full QCD as effects of
an altered running of {\em the} strong coupling constant with
{\em one} relevant scale is an overly simplistic
picture. We conclude that neglection of sea quarks results
in a scatter between different scale determinations 
of up to 20~\%, a systematic uncertainty
of the quenched approximation that cannot be {\em repaired}
or discussed away.

Present day high precision data on the (partially) unquenched
static QCD potential
is limited to two light quarks with
mass $m_q>25$~MeV.
Apart from one very exploratory study~\cite{mtc},
determinations of relativistic corrections for dynamical fermions
are completely missing.
Within statistical accuracy, data on the static potential
obtained with two flavours of Wilson sea quarks as well as within
the quenched approximation
are compatible with the parametrisation,
$V_0(r) = \kappa r-e/r$,
down to distances as small as $0.2$~fm.
{}From a fit to data with $r\geq r_{\min}=0.2$~fm
one of the authors has obtained $e= 0.292(5)$
in quenched simulations and 
$e= 0.352(9)$ in the
two flavour case with $m_u=m_d\approx 50$~MeV~\cite{sesaminprep}.
If we allow for an additional
term, $f/r^2$, to mimic running coupling effects, we end up with
$e\approx 0.32$ and $e\approx 0.37$, respectively.
The latter fits incorporate the running of the Coulomb coefficient to
distances $r\approx 0.1$~fm, significantly less than the mean
radius of $\Upsilon(1S)$, $\langle r^2\rangle^{1/2}\approx 0.25$~fm,
but only affect the parameter $e$ slightly, in
comparison to the
change induced by ``switching on'' two sea quarks.
We conclude that the renormalisation of the effective
intermediate distance Coulomb-coefficient
is the dominant effect of unquenching, rather than the
slightly different running of the coupling with the scale at large
momenta.

For the sake of consistency with Ref.~\cite{baliwachter}
we use the value $e=0.32$  throughout this article
for what we call ``quenched'' while we estimate a value
$e\approx 0.40$ for what we consider to be the real world
with three flavours of light quarks. 
It is apparent from both, lattice data on the $1P\,1S$ splitting,
and from the above information on the static potential
that the Coulomb term is indeed underestimated
in the quenched approximation. In what follows, we will investigate
the dominant effect of unquenching by just varying the Coulomb 
coefficient in a Cornell potential simulation.

\section{Method}
\label{Method}
Throughout this paper we only use the Cornell-type
parametrisation of QCD potentials, and neglect running coupling effects
which would weaken the interaction at short distance.
This means that we are likely to (slightly) overestimate fine structure
splittings
while the effect on spin averaged splittings is small. Physical states with
small spatial extent like the $1S$ state should be thought of as
being shifted into
the direction of the corresponding prediction at somewhat weaker
Coulomb coupling.
Within this model, we estimate the sizes of various
systematic effects, allowing conclusions on the degree
of accuracy of lattice predictions on the $\Upsilon$ system.

Quenching effects and the importance of relativistic corrections are studied
by solving a continuum Schr\"odinger equation by use of the Numerov method.
Like in Ref.~\cite{baliwachter}, we start from the Hamiltonian,
\begin{equation}
H = 2\left[m + \delta m(m)\right] + H_0 + \delta H_{\mbox{\scriptsize kin}}
+ \delta H_{\mbox{\scriptsize SI}}
+\delta H_{\mbox{\scriptsize SD}},
\end{equation}
with
\begin{equation}
H_0=\frac{{\mathbf p}^2}{m}+\overline{V}(r),
\end{equation}
where
\begin{equation}
\label{potcent}
\overline{V}(r)=V_0(r)-\frac{1}{4m^2}\left\{c_D(m)b-
2\left[c_D(m)-\frac{1}{3}\right]\kappa\right\}\frac{1}{r}.
\end{equation}
The parametrisation,
\begin{equation}
\label{potstat}
V_0(r) = \kappa r-\frac{e}{r},
\end{equation}
corresponds to the static QCD potential, the
additional terms of Eq.~(\ref{potcent})
giving rise to order $v^4$ corrections to the energy levels.
After solving the unperturbed Schr\"odinger equation,
\begin{equation}
\left(H_0-E_{nl}\right)\psi_{nll_z}({\mathbf x})=0,\quad
\psi_{nll_z}({\mathbf x})=\frac{u_{nl}(r)}{r}Y_{ll_z}(\theta,\phi),
\end{equation}
the remaining order $v^4$ relativistic corrections are computed as
perturbations. 
The kinetic correction reads as follows,
\begin{equation}
\delta H_{\mbox{\scriptsize kin}}=-\frac{p^4}{4m^3}.
\end{equation}
For the spin independent corrections we obtain,
\begin{equation}\label{potsi}
\delta H_{\mbox{\scriptsize SI}}=\frac{1}{2m^2}
\left\{4\pi e\left[1+\frac{c_D(m)}{2}-d_s(m)\right]\delta^3(r)
-\frac{2e}{r}p^2-\left(\frac{\kappa}{3r}-\frac{e}{r^3}\right)L^2\right\},
\end{equation}
while the spin dependent correction terms read,
\begin{eqnarray}
\delta H_{\mbox{\scriptsize SD}}(r)&=&\frac{1}{2m^2}\left\{
\left[-\frac{\kappa}{r}+ \frac{4c_F(m) (e-h) -e}{r^3}\right]
{\mathbf L\cdot S}\right.\nonumber\\\label{potsd}
&+&6c_F^2(m)\frac{(e-h)}{r^3}T\\\nonumber
&+&\left.8\pi\left[2c_F^2(m)-d_v(m)\right](e-h)
\delta^3(r)\frac{\mathbf
S_1\cdot S_2}{3}\right\},
\end{eqnarray}
with
\begin{eqnarray}
\frac{{\mathbf S_1\cdot S_2}}{3}&=&
\frac{1}{6}\left[s(s+1)-\frac{3}{2}\right],\\
\mathbf L\cdot S
&=&\frac{1}{2}\left[j(j+1)-l(l+1)-s(s+1)\right],\\
 T=
\frac{{\mathbf x\cdot S_1}{\mathbf x\cdot S_2}}{r^2}
-\frac{\mathbf S_1\cdot S_2}{3}&=&
-\frac{6({\mathbf L\cdot S})^2+3{\mathbf L\cdot S}-2s(s+1)l(l+1)}
{6(2l-1)(2l+3)}.
\end{eqnarray}
In the present investigation, we choose to 
approximate the matching coefficients by their tree-level
values,
$c_F(m)=c_D(m)=1$ and $d_s(m)=d_v(m)=\delta m(m)=0$,
as it is usually done in
Lattice NRQCD simulations too.
If the quark mass $m$ differs from the gluon momentum cut-off
$\mu$ at which the potentials have been evaluated, radiative
corrections have to be considered\footnote{Results for the one-loop running of $c_F(m)$ can be found in
Ref.~\cite{eichhill}, for $c_D(m)$ in Refs.~\cite{balzereit,manohar}
and for $d_v(m)$ in Ref.~\cite{chen}.
A two-loop renormalisation group improved
result for $c_F(m)$ has been obtained recently~\cite{amoros}.
The binding energy of the $\Upsilon$ ground state is known to order
$m\alpha_s^4$ in perturbation theory~\cite{pineda}.
The Hamiltonian for the unequal quark mass case
can be found in Refs.~\cite{baliwachter,antonio}. $d_s(m)$ can be
obtained from Ref.~\cite{soto}.}.

With relations, such as $\langle nll_z| f(r){\mathbf p}^2|nll_z\rangle
=m\langle nll_z|f(r)[E_{nl}-\overline{V}(r)]|nll_z\rangle$, all perturbations
can be reduced to functions of the expectation values
$\langle r^{\alpha}\rangle$ with $\alpha=-4,\ldots,1$
and $4\pi\langle \delta^3(r)\rangle=|\psi(0)|^2$ and
the unperturbed
energy levels $E_{nl}$.
We use the parameter values of Ref.~\cite{baliwachter},
\begin{equation}
\label{eq:params}
r_0^{-1}=406 \mbox{~MeV},\quad h = 0.065,\quad b =3.81\kappa.
\end{equation}
We can convert $r_0$, defined as the distance $r$ at which~\cite{sommer}
$r^2dV(r)/dr=1.65$, into the string tension,
\begin{equation}
\label{eq:string}
\kappa=\frac{1.65-e}{r_0^2}.
\end{equation}
The above
value of $r_0$ has been obtained by optimising the spectrum with respect to
all experimentally observed bottomonium states. The advantage
of this procedure over a matching of just the two lowest lying
splittings is that this $r_0$ comes out to be in closer
agreement with scales obtained from light hadronic observables, such as
$m_{\rho}$ or $f_{\pi}$. $e$ is varied starting from the
(quenched) value $e=0.32$ in order to estimate sea quark effects while
$m$ is varied between 1 and 15~GeV for investigation of the mass dependence
of splittings. The values of $m$ that optimally reproduce the bottomonium
and charmonium levels (within the quenched set-up) have been found to be
$m_b\approx 4.68$~GeV and $m_c\approx 1.33$~GeV, respectively. An increase
of $e$ by 25~\% results in an increase of these mass estimates
by only 1~\% and 4~\%, respectively.

We expect 
electromagnetic interactions to yield an
increase of the parameter
$e$ by $(1/3)^2\alpha_f(m_b)\approx 10^{-3}$ for
$\Upsilon$ states and $(2/3)^2\alpha_f(m_c)\approx 3.5\times 10^{-3}$
for $J/\psi$ states.
thereby having a negligible impact on the spectrum.
$\alpha_f$ denotes the QED fine structure constant. The change
results in an increase of about 2~MeV for a QCD prediction of $m_b$ and 
of about 8~MeV for $m_c$.

We estimate FSE
for the $\Upsilon$ system by numerically solving the Schr\"odinger equation
on three-dimensional cubic lattices. For this purpose, we work at
order $v^2$ only and employ the
following definition of the
lattice Laplacian,
\begin{equation}
\nabla^2 \psi({\mathbf x}) = \frac{1}{a^2}
\sum_{j=1}^3\left[\psi({\mathbf x} + a{\mathbf e}_j)+
\psi({\mathbf x} - a{\mathbf e}_j) - 2\psi({\mathbf x})\right],
\end{equation}
which is correct up to order $a^2$ lattice artifacts
(${\mathbf e}_j$ denotes a unit vector in direction
$j$). Note that --- to this order in $v$ --- we use the
standard Schr\"odinger equation,
\begin{equation}
\left[\frac{p^2}{m}+V_0(r)-E_{n}\right]\psi_{n}({\mathbf x})=0,
\end{equation}
with the static potential $V_0(r)$ of Eq.~(\ref{potstat}) only,
instead of $\overline{V}(r)$ of Eq.~(\ref{potcent}). $n$ is to be understood
as a multi-index that completely characterises a given state.
We choose the coordinate system such that the origin is in the
centre of an elementary cube, therefore avoiding problems 
with the divergence of $V_0(r)$ at $r=0$.
Throughout the simulations the parameters of Ref.~\cite{baliwachter}
are used:
\begin{equation}
\sqrt{\kappa} = 468 \mbox{~MeV},\quad e = 0.32,\quad m=4.676 \mbox{~GeV}.
\end{equation}

The discretised Hamiltonian is solved by use of
a successive over-relaxation algorithm.
Excitations are accessed via Gram-Schmidt orthogonalisation
with respect to previously obtained solutions. Only one octant
needs to be simulated, and different 
orbital states are generated by enforcing combinations of periodic (even)
or anti-periodic (odd) boundary conditions along various lattice
directions.
We use the $z$-axis as our quantisation axis. Thus, we always
apply the same boundary conditions along the $x$- and $y$-directions,
such that four different combinations are realised:
even $x$, even $z$ ($ee$); even $x$, odd $z$ ($eo$); odd $x$, even
$z$ ($oe$); odd $x$, odd $z$ ($oo$). All other solutions can be obtained
as linear superpositions of the solutions obtained in this manner
with eventually permuted
lattice axes.

For a cubic lattice the relevant discrete symmetry group is $O_h$.
States can be classified in accord to the five irreducible
representations $A_1$, $A_2$, $E$, $T_1$ and $T_2$. $A_1$ and
$A_2$ are one-dimensional, $E$ is two-dimensional and $T_1$ and
$T_2$ both are three-dimensional.
The implemented boundary conditions determine the symmetry of the
wave function under reflections with respect to a lattice plane through
the origin.
While this suffices to unambiguously
single out $T_1$ ($eo$), $T_2$ ($oe$) and
$A_2$ ($oo$) states, both $A_1$ and $E$ states fall into the same $ee$
class.

Each $O_h$ state has overlap to various continuum states with
different angular momenta.
In order to allow for an identification
of the continuum spin content, the expectation value $\langle L^2\rangle =
l(l+1)$
has been traced. Moreover, lattice spacing and volume have been
changed and continuum and infinite volume extrapolations performed.
In Table~\ref{tab:rep}, we have listed which states we expect to find within
each set of boundary conditions. $S$ waves can only be
found with $ee$ boundary conditions and $P$ waves only with
$eo$ boundaries. In the latter case, with antiperiodic boundaries in the
$z$-direction, we will only obtain the $l_z=0$ state.
$D$ waves will be generated with both,
$ee$ boundaries (two states, corresponding
to $l_z=\pm  2$) and with $oe$ boundaries (three states of which only the
$l_z=0$ state is possible with odd boundary conditions in the $z$ direction).
Finally, $F$ waves will be obtained in the
$eo$ and $oe$ (three states in each sector, of which we will only find
one in either case) as well as in the $oo$ (one state) sectors.

Lattices with octants of
up to $72^3$ sites
are being realised and lattice spacings down to
$a=0.025$~fm are simulated, enabling us to study both, finite volume effects
and the restoration of rotational symmetry
between different lattice representations with overlap to the same continuum
angular momentum.

\section{Results}
\label{Result}
\subsection{Relativistic and radiative corrections}
\label{correc}

\begin{figure}[htp]
\epsfxsize=11truecm
\centerline{\epsffile{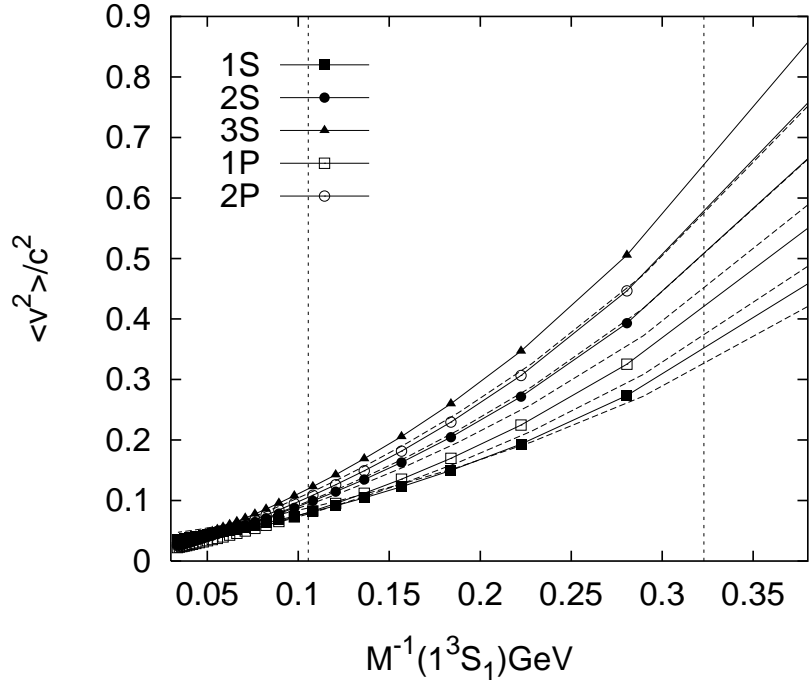}}
\caption{Average quark velocities for various states versus the inverse
$1^3S_1$ mass. Solid curves correspond to $e=0.32$, dashed curves to
$e=0.40$.}
 \label{fig3}
\end{figure}

We intend to estimate the uncertainties on spin averaged
order $v^4$ NRQCD level splitting predictions, due to higher order
corrections.
For this purpose, in Fig.~\ref{fig3}, we have plotted the average velocity
$\langle nl|v^2|nl\rangle$ for various states as a function of
the inverse quarkonium ground state mass, $M_{\Upsilon}^{-1}$.
The vertical lines correspond to bottomonium and charmonium. For
bottomonia states, we typically find $\langle v^2\rangle
\approx 0.1$
while for charmonia we obtain $\langle v^2\rangle\approx 0.4$.
Two sources of uncertainties that are common to both, the
potential approach and Lattice NRQCD, exist:
(a) radiative corrections to the coefficients of the NRQCD
Lagrangian and (b) higher order relativistic corrections.
Note that for the moment being our estimates apply
to an exact prediction from
order $v^4$ NRQCD with tree-level matching coefficients.
We ignore errors that are
due to the lattice discretisation or an inadequate representation
of the potentials at short distance by the parametrisation used.
In particular for $S$ wave
and, to a lesser extent, $P$ wave fine structure splittings
we expect such effects to be another significant source of uncertainty.

If we start from the non-relativistic result for a mass splitting,
$\Delta M_{v^2}$, calculated at order $v^2$, the correct relativistic
splitting can be obtained as an
expansion in powers of the typical heavy quark velocity $v^2$,
\begin{equation}
\Delta M = \Delta M_{v^2}\left(1+c_1v^2+c_2v^4+\cdots\right).
\end{equation}
We find values $c_1v^2=0.034(0.036)$ for the
$\overline{2S}\,\overline{1S}$ and $c_1v^2=0.025(0.029)$ for
the $\overline{1P}-\overline{1S}$ bottomonium splittings.
The first numbers refer to the quenched estimate ($e=0.32$) while the
numbers in brackets refer to $e=0.40$. The {\em overline} symbol denotes
arithmetic
averaging of all masses of states with the given quantum numbers;
apparently, the sizes of the correction
terms only weakly depend on quenching.
For the $J/\psi$ system, we obtain
the values $0.103(0.115)$ and $0.070(0.082)$, respectively.
Thus, in both cases we find, $c_1\approx 1/3$. 

We have set all coefficients of the NRQCD Lagrangian
to their tree-level values. However, when matching the effective 
theory to QCD, radiative corrections come into play.
As long as we are only concerned with the spin averaged spectrum,
$\delta H_{\mbox{\scriptsize SD}}$ [Eq.~(\ref{potsd})] does not contribute.
Therefore, the only uncertainties enter in the terms
of Eqs.~(\ref{potcent}) and (\ref{potsi})
that contain the coefficient $c_D(m)$.
Discarding these terms
results in a shift of
the $\overline{2S}\,\overline{1S}$ splitting of 5~MeV
and 16~MeV in the $\overline{1P}\,\overline{1S}$ splitting
for bottomonia and 2~MeV and 16~MeV
for charmonia states, respectively.
Based on the
one-loop results of Refs.~\cite{balzereit,manohar}, we estimate the uncertainty
in $c_D$ to be as big as 25~\% for $\Upsilon$ states and 100~\%
for $J/\psi$ states\footnote{In Ref.~\cite{baliwachter},
one of the present authors has quoted somewhat
different numbers that were based on the calculations
of Ref.~\cite{chen}. However, an error in the re-parametrisation
invariance relations used in this reference has recently been
discovered~\cite{georgi}.} for simulations on lattices with
resolutions $a\approx 0.08$~fm.

By assuming $c_2<c_1$, which is motivated by the observation that
$c_1\approx 1/3<c_0=1$, we expect the order $v^6$ correction
to be smaller than 2~MeV and 1.3~MeV for bottomonia
$\overline{2S}\,\overline{1S}$ and
$\overline{1P}\,\overline{1S}$ splittings, respectively. In case of charmonia,
we can only roughly estimate the corresponding values
to be 45~MeV and 20~MeV.
Taking the additional uncertainty in $c_1$ due to radiative corrections
to $c_D$ into account, we expect to be accurate within 5--6~MeV for $\Upsilon$
$\overline{1P}\,\overline{1S}$ and 3--4~MeV for $\overline{2S}\,\overline{1S}$
splittings which corresponds to a
relative precision of about 1~\%, while of the $J/\psi$
system we estimate 50~MeV and 40~MeV, respectively, which amounts to
an expected accuracy of only 10~\%.

Unfortunately, the singlet $S$ states for the $\Upsilon$ system
have not been discovered in experiment yet. The triplet levels are
increased by 10(12)~MeV and 6.5(7.5)~MeV in respect to the corresponding
spin-averaged results, obtained at $e=0.32 (0.40)$,
for $1S$ and $2S$ states, respectively.
By allowing
an uncertainty of 25~\% on the matching coefficient
within Eq.~(\ref{potsd}), we estimate cumulated uncertainties of
about 8--9~MeV and 5--6~MeV for bottomonium $\overline{1P}1^3S_1$
and $2^3S_11^3S_1$ splittings, respectively.
Note that a further uncertainty of up to
10~MeV has to be added to
lattice results on the $1^3S_1$ state obtained
around $a\approx 0.08$~fm to account for discretisation
errors of the wave function at the origin~\cite{inprep2}.

The uncertainty in the $P$ wave
fine structure, which by itself is an order $v^4$
effect, is much bigger of course. One might hope --- again by
assuming $c_3<c_2$ --- that
relativistic corrections are suppressed by a factor of order
$v^2$ in respect to the leading order result. This alone would
imply an uncertainty of 10~\% for the bottomonium and 40~\% for
the charmonium fine structure. The uncertainty in radiative
corrections is dominated by the error of $c_F^2(m)$,
which, based on the two-loop calculation of Ref.~\cite{amoros},
we expect to be about 25~\% for bottom quarks and
70~\% for charm quarks. In linearly adding these
two sources of error, we end up with relative precisions
of 35~\% and 110~\% for order $v^4$ $\Upsilon$ and $J/\psi$
fine structure splitting predictions, respectively, i.e.\ the predictive power
for the charmonium fine structure is {\em zero}.
However, meaningful results can still be obtained for ratios of
fine structure splittings since these turn out to be (almost)
independent of
the matching coefficients.

\subsection{Mass dependence and quenching effects}

\begin{figure}[htp]
\epsfxsize=11truecm
\centerline{\epsffile{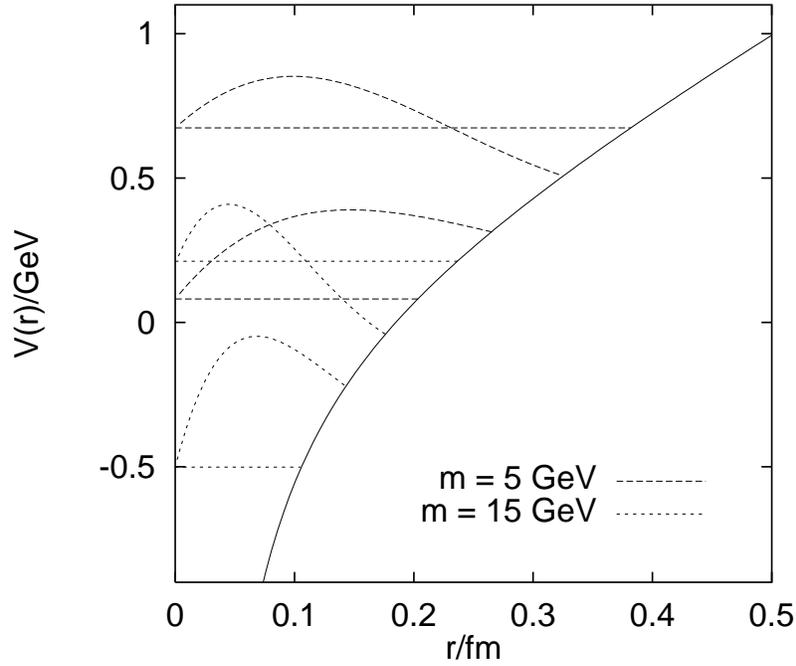}}
\caption{$2S$ and $1S$ levels and wave functions in comparison with
the inter-quark potential ($e = 0.4$, $r_0^{-1}=406$~MeV) for two
quark masses.}
 \label{fig1}
\end{figure}

\begin{figure}[htp]
\epsfxsize=11truecm
\centerline{\epsffile{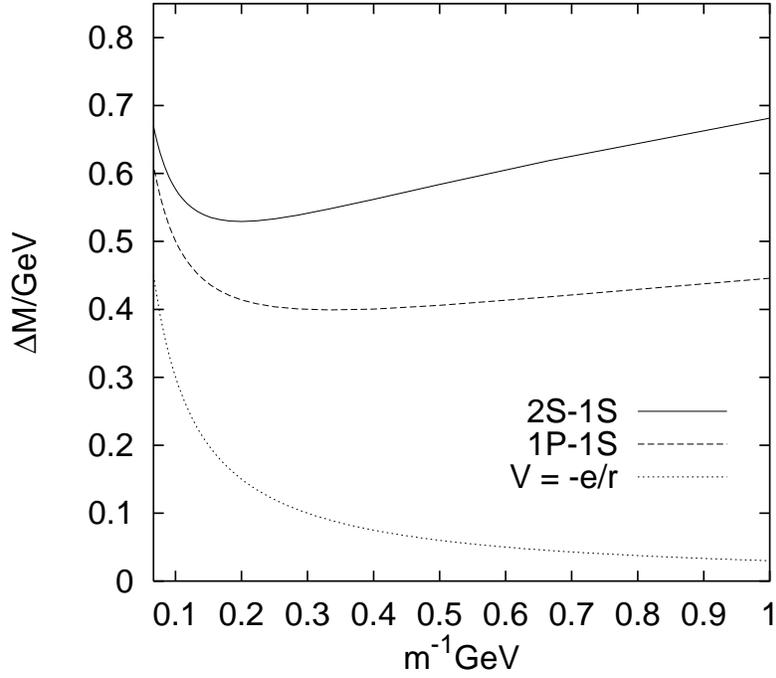}}
\caption{Spin averaged $2S\,1S$ and $1P\,1S$ splittings versus the inverse
quark mass ($e = 0.4$).}
 \label{fig2}
\end{figure}

As has been argued above, when the quark mass is increased quarkonia
states move deeper into the Coulomb region and, with $m$ being
the only remaining dimensionful parameter, all splittings will
eventually diverge linearly with the quark mass.
This can be seen from Fig.~\ref{fig1} where we compare
the $2S\,1S$ splitting (calculated at order $v^2$) 
at a quark mass $m=5$~GeV with that
at $m = 15$~GeV. Of course, this also means that harder and harder
gluons will obtain momenta lower than the (diverging) ultra-soft scale,
$mv^2$, and the potential approach will loose validity, due to
significant retardation effects. If we ignore this problem for the moment
being, for very heavy quark mass we would expect the $2S\,1S$ and
$1P\,1S$ splittings to become degenerate and
diverge like,
\begin{equation}
M(2S)-M(1S)=\frac{3}{16}e^2m.
\end{equation}
The situation is visualised in Fig.~\ref{fig2} for the unquenched case.
The dotted curve is the expectation for a pure Coulomb potential
which will asymptotically be approached by the two other curves as
$m$ is increased.
Around the bottom quark mass this pure Coulomb contribution amounts to about
20~\% of the splittings while for the charm quark mass it contributes only
slightly more than 5~\%.
A similar divergent behaviour is expected to set in for the
$3S\,1S$ and $2P\,1S$ splittings at somewhat bigger quark masses.

\begin{figure}[htp]
\epsfxsize=11truecm
\centerline{\epsffile{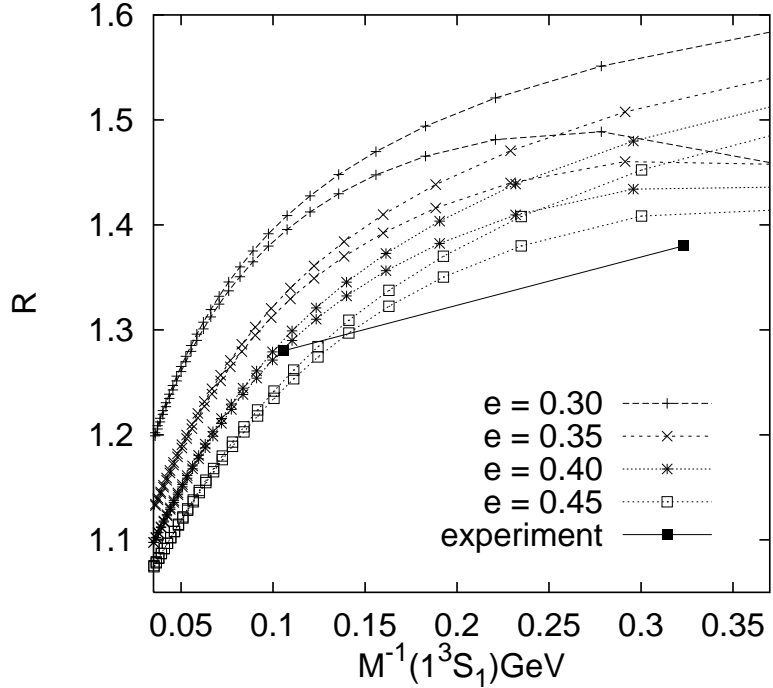}}
\caption{
$R=\Delta M_{2S\,1S}/\Delta M_{1P\,1S}$ as a function of
$M_\Upsilon^{-1}$ for various values of the Coulomb coupling.
The upper curves correspond to the static potential, the lower ones
incorporate relativistic corrections.}
 \label{mass1}
\end{figure}

In Fig.~\ref{mass1}, we display the dependence of the ratio,
\begin{equation}
R=\frac{M(2^3S_1)-M(1^3S_1)}{M(\overline{1^3P})-M(1^3S_1)},
\end{equation}
on the inverse quarkonium ground state
mass for various values of the Coulomb strength.
For this purpose, we operate at fixed $r_0=406$~MeV [Eq.~(\ref{eq:params})]
and adjust the string tension in accord with Eq.~(\ref{eq:string}).
The upper curves correspond to the order $v^2$ results, the lower curves
to the order $v^4$ predictions.
The Figure suggests a value $e\approx 0.40$ for real world QCD,
which is somewhat larger than the $n_f=2$, $m_u=m_d\approx 50$~MeV upper
limit $e=0.37(1)$ from a fit to a parametrisation that incorporates
an $f/r^2$ term.
Given the observed quark mass and flavour
dependence~\cite{sesaminprep,cppacs},
the value $e\approx 0.40$ appears to be very reasonable.
Within the systematic uncertainties discussed above, the
experimental value of $R$ for the $J/\psi$ system
is compatible with this choice of $e$ too.
We obtain a quenched ($e=0.32$)  bottomonium estimate,
$R\approx 1.38$ which, while agreeing
with Lattice NRQCD results~\cite{nrqcd2}, disagrees
with experiment. The failure of
the quenched potential to reproduce the observed
$\Upsilon$ spectrum
has also been noticed in Ref.~\cite{mich}.

\begin{figure}[htp]
\epsfxsize=11truecm
\centerline{\epsffile{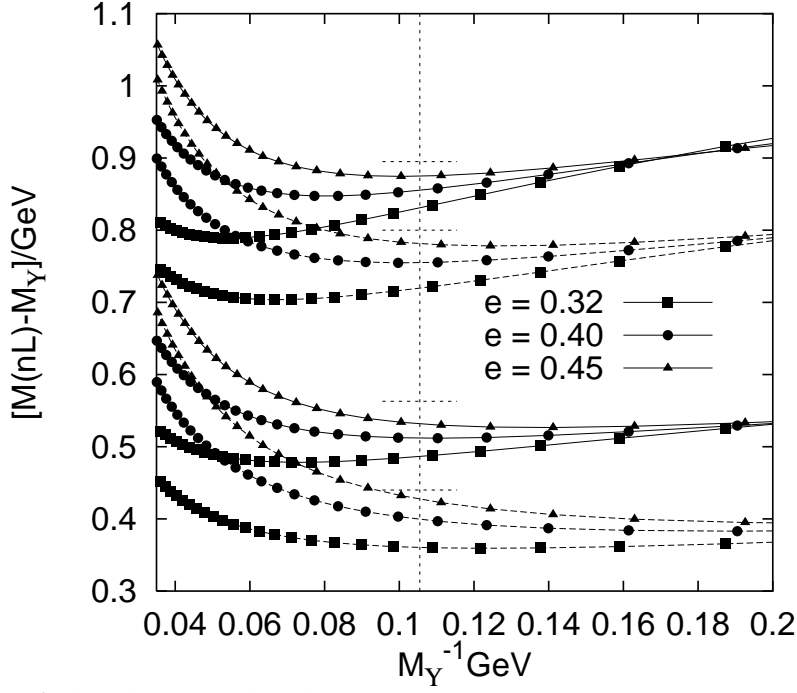}}
\caption{Order $v^4$ $2^3S_11^3S_1$ and $3^3S_11^3S_1$ splittings
(solid curves) as well as
$\overline{1P}1^3S_1$ and $\overline{2P}1^3S_1$
splittings (dashed curves) versus
the inverse $\Upsilon$ mass for various values of the Coulomb coupling.
The dashed horizontal and vertical lines denote the experimental
values.}
 \label{mass2}
\end{figure}

In Fig.~\ref{mass2}, we display the splittings between
the $1^3S_1$ $\Upsilon$ ground state and $2^3S_1$ and $3^3S_1$ states
(solid curves) as well as
$\overline{1^3P}$ and $\overline{2^3P}$ states (dashed curves)
for three values of the Coulomb coefficient, calculated to order $v^4$.
All splittings decrease
slightly with increasing quarkonium mass until they reach a
turning point, after which they start to diverge. The position of
this point critically depends on the level;
for physically larger states it is shifted towards
higher masses. It also depends
on the strength of the Coulomb coefficient which explains why
$R$ reacts in a very sensitive way towards quenching.
While most level differences can be reasonably well reproduced with
values $0.40\leq e\leq 0.45$, all splittings with respect to
$1^3S_1$ are considerably underestimated, an effect that
cannot be absorbed into redefinitions of the quark mass
$m_b$ and the scale $r_0$ alone. We also note that differences between
excitations and the $2^3S_1$ or $\overline{1P}$ levels exhibit
a reduced dependence on $e$, compared to the
splittings with respect to the physically smaller $1^3S_1$ state that are
displayed in the figure.

By incorporating a weakening of the QCD coupling at shorter
distance into the parametrisation of the interaction
potential~\cite{eichq},
the prediction can be brought in line
with experiment within the estimated uncertainties of
Sec.~\ref{correc}. The $1S$ state turns out to be
most sensitive towards the running of the coupling.
In quenched Lattice NRQCD simulations~\cite{nrqcd2},
after adjusting the scale
from the $2^3S_11^3S_1$ and $\overline{1P}\,1^3S_1$ splittings,
all other level spacings have been found to be overestimated,
and this despite the fact that
our study of FSE suggests that the NRQCD $3^3S_1$ mass is underestimated
by up to 50~MeV on the finite simulation volume.
The potential results suggest that the
disagreement within the $\Upsilon$ spectrum
as well as between determinations of the lattice spacing
from light hadronic quantities and quarkonia level splittings will
indeed be reduced in unquenched simulations.
A determination of the minimal lattice resolution required to 
resolve the running
of the QCD coupling sufficiently well for a precision determination of
$\Upsilon$ levels is the subject of an ongoing study~\cite{inprep2}.

\begin{figure}[htp]
\epsfxsize=11truecm
\centerline{\epsffile{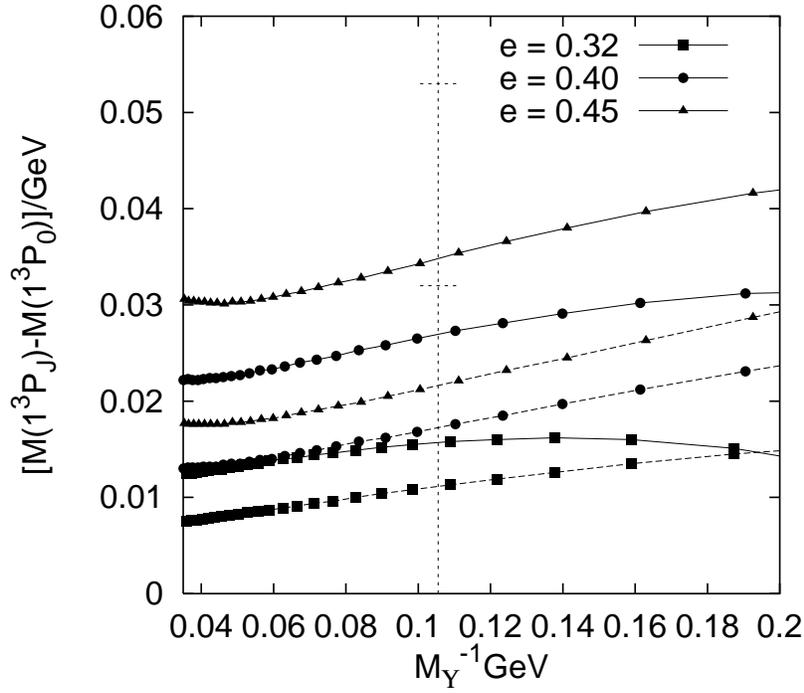}}
\caption{The $1^3P_j$ fine structure as a function of the inverse
$\Upsilon$ mass for various values of the Coulomb coupling.
Solid curves denote the difference $M(1^3P_2)-M(1^3P_0)$, dashed
curves $M(1^3P_1)-M(1^3P_0)$.
The dashed horizontal and vertical lines denote the experimental
values.}
 \label{mass3}
\end{figure}

In Fig.~\ref{mass3}, we investigate the $1^3P_2\,1^3P_0$ (solid
curves) and $1^3P_1\,1^3P_0$ (dashed curves) fine splittings
as a function of the inverse $\Upsilon$ mass for three values of
the Coulomb coefficient. For $e=0.40$, we underestimate
experiment by a factor of almost {\em two}.
If we had included a running coupling
into our parametrisation, the predicted splittings might have been
even somewhat smaller.
For lattice spacings 0.067~fm$ <a<0.092$~fm, by assuming (somewhat
arbitrarily) an ${\overline{MS}}$-scheme gluon
momentum cut-off $\mu=\pi/a$,
we obtain a one-loop estimate $1.06<c_F^2(m)<1.13$~\cite{baliwachter},
a range far away from the required 80--90~\% correction. This result
indicates, if we accept the conjecture that the spectrum
should be described by QCD, that higher order contributions to
the matching coefficients are important.
As expected, predictive power
of the potential approach (and Lattice NRQCD) on the fine structure
is indeed limited by a huge systematic
uncertainty. Note that the
difference has nothing to do with uncertainties in ``tadpole'' factors
since a non-perturbative renormalisation of
lattice results with respect to the continuum is included into the
parametrisations published in Ref.~\cite{baliwachter}.
We expect to be accurate within 10~\%
for ratios of fine structure splittings, though.
Indeed, the ratio
of the two splittings at $e=0.40$ comes out to be
$\rho\approx 1.56$, compared to the experimental value $\rho=1.66$.

\subsection{Finite size effects}

\begin{figure}[htp]
\epsfxsize=11truecm
\centerline{\epsffile{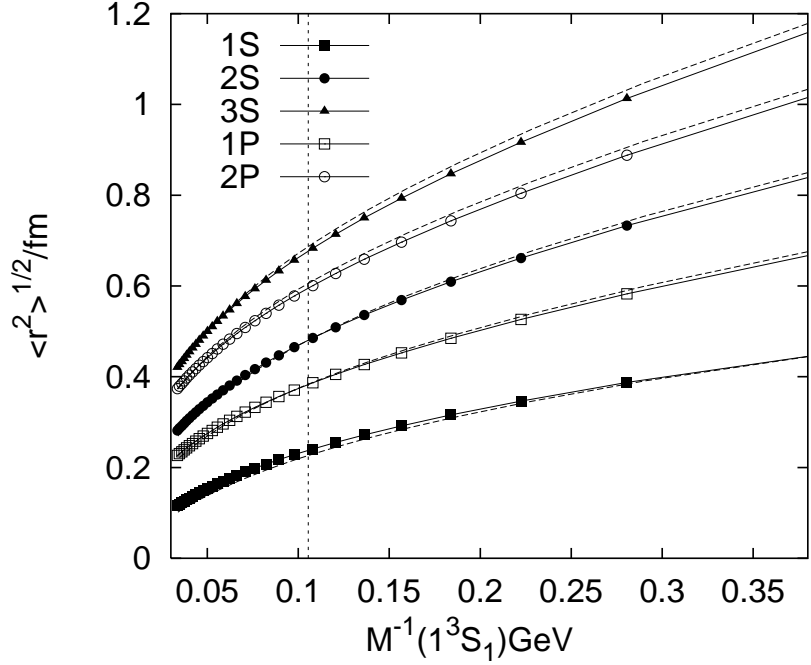}}
\caption{Radius of wave functions versus the inverse $\Upsilon$
mass. Solid curves correspond to $e=0.32$, dashed curves to $e=0.40$.}
 \label{rad}
\end{figure}

In this Section, we will investigate FSE on
the spin averaged spectrum for the quenched case. For this
purpose, we restrict ourselves to the order $v^2$ Hamiltonian.
In Fig.~\ref{rad}, we display the root mean squared
radii of various wave functions for $e=0.32$ and $e=0.40$. Since the spatial
extents of the states
are apparently only weakly affected by the change in $e$,
FSE in unquenched simulations are likely to behave in a very similar way,
up to string breaking effects.
FSE on the fine structure are expected to be negligible due to the short range
nature of spin interactions\footnote{Only the ${\mathbf L\cdot S}$
term of Eq.~(\ref{potsd}) contains a long range $1/r$
interaction. This contribution, however,
is numerically tiny.}. Note that all effects that we will discuss
are entirely due to squeezing of the wave
function and the presence of mirror charges on a torus.
In lattice simulations additional sources of
FSE in general exist and the behaviour of the
quark propagator itself might be affected by the volume.
Our prejudice, however, is that for heavy quarks such an effect
can be neglected, as long as one remains comfortably separated from the
deconfinement transition. This is definitely the case
on volumes of $(1.5$~fm$)^3$. 

\begin{figure}[htp]
\epsfxsize=11truecm
\centerline{\epsffile{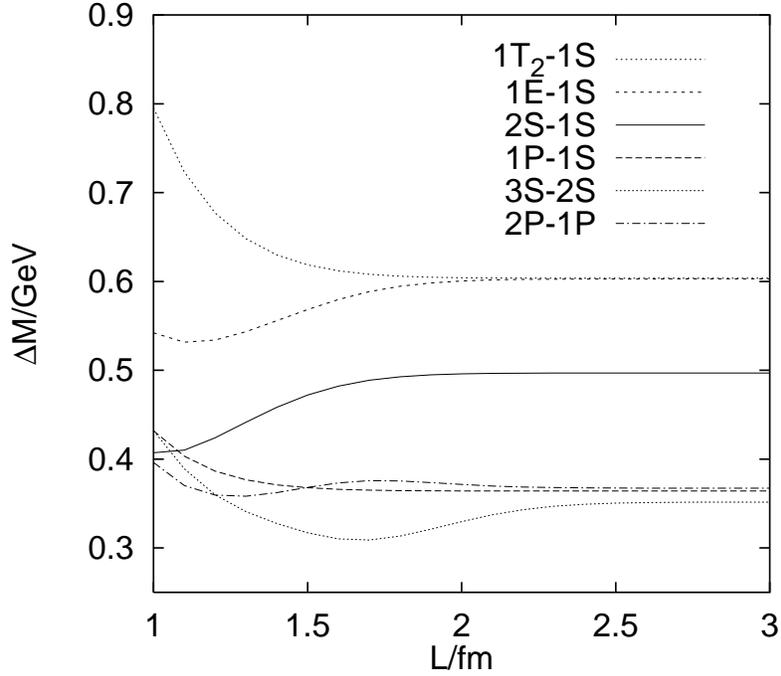}}
\caption{$\Upsilon$ splittings as a
function of the lattice extent $L$.}
 \label{fse1}
\end{figure}

\begin{figure}[htp]
\epsfxsize=11truecm
\centerline{\epsffile{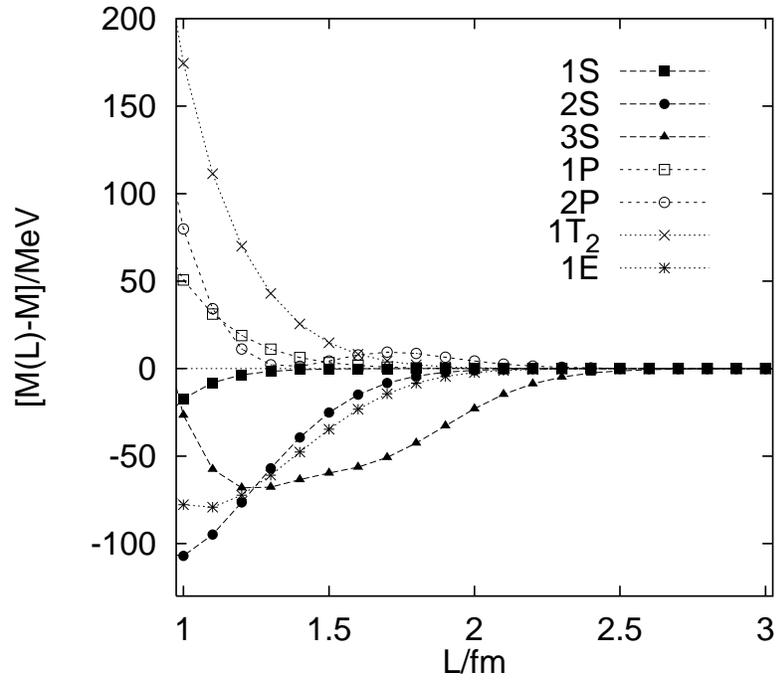}}
\caption{Difference between $\Upsilon$ levels and their corresponding
infinite volume values as a function of the lattice extent $L$.}
 \label{fse2}
\end{figure}

In Fig.~\ref{fse1}, we investigate the effect of a finite volume onto
various splittings. The results have been obtained
with lattice spacings $a=0.025, 0.05,0.1$ and 0.2~fm and are extrapolated 
to the continuum limit.
Note that on the cubic lattice, the $1D$ state exists in
two representations, $1T_2$ and $1E$, that will only become degenerate
in the infinite volume continuum limit.
Obviously,
the spectrum is very much affected by the box size for a
lattice extent $L<1.5$~fm
while for $L>2$~fm most levels have effectively approached their
infinite volume limits.
However, for the $3S$ level an extent as
big as $2.5$~fm is required. 
Another representation of the data is displayed in Fig.~\ref{fse2}
where we show $\delta M(L)=M(L)-M(\infty)$ as
a function of $L$.
In Table~\ref{tab:fse} we display the minimal lattice extents
$L_{\mbox{\scriptsize 3 MeV}}$ at which FSE were found to be
smaller than 3~MeV for various states. For this purpose, the box size has been 
increased in steps of 0.05~fm.
We have also included the root mean squared radii of Fig.~\ref{rad}.
We find a linear behaviour,
$L_{\mbox{\scriptsize 3 MeV}}=a\langle r^2\rangle^{1/2}+b$
with $a\approx 2.25$ and $b\approx 0.65$~fm.

The finite size behaviour of the $1S$ level
is well parameterised for $L>1$~fm by a polynomial in $L^{-3}$,
\begin{equation}
\delta M(L) = \sum_{i=1}^4\left(\frac{a_i}{L}\right)^{3i},
\end{equation}
with
\begin{equation}
a_1=-0.088 \mbox{~fm},\quad
a_2= 0.477 \mbox{~fm},\quad
a_3=-0.712 \mbox{~fm},\quad
a_4= 0.717 \mbox{~fm}.
\end{equation}
The leading order coefficient is found to be rather small,
compared to those of higher order correction, which is consistent
with the fact that FSE set in rather suddenly.
Parametrisations for other states are consistent with small
coefficients of $L^{-3}$ terms
too. While the finite size behaviour for $n=1$ states
is monotonous, this is not the case for radial excitations anymore.
In particular, the result for $3S$ (triangles) suggests
that for some states infinite volume extrapolations
of finite volume data might become dangerously non-trivial.

\begin{figure}[htp]
\epsfxsize=14truecm
\epsffile{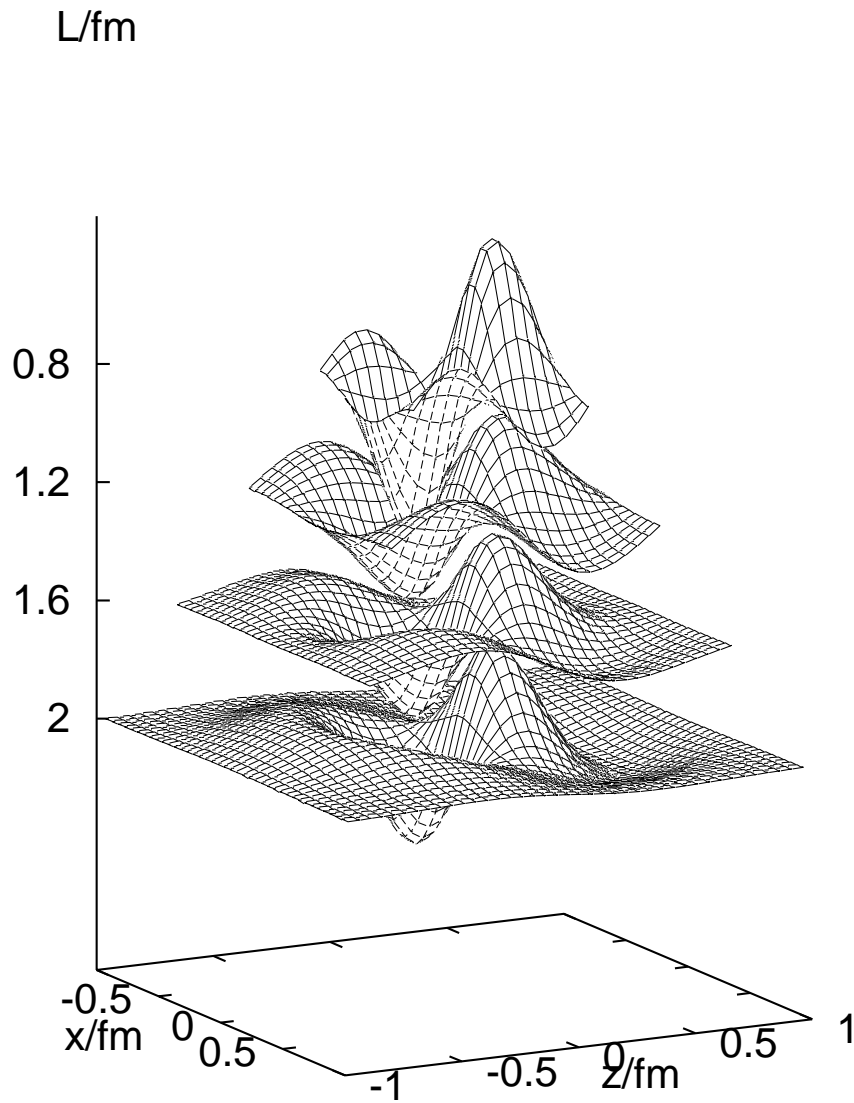}
\caption{The $2P$ wave function at $y=0$ for various
lattice volumes.}
 \label{fse3}
\end{figure}

In Fig.~\ref{fse3} we display the $2P$ wave function, obtained
on an $a=0.05$~fm lattice for various lattice extents, varying from
0.8 up to 2~fm. We show a cross section through the $x-z$ plane at
$y=0$. Of course, the wave function vanishes within the plane $z=0$ and
exhibits approximate rotational symmetry around the $z$-axis.
{}From Fig.~\ref{fse2}
one sees that the energy decreases up to 1.3~fm and increases
again thereafter, until a second turning point is approached
around 1.7~fm. From Fig.~\ref{fse3} we conclude that the second node
of the wave function only fits onto the lattice for an extent slightly
bigger than 1.2 fm, which explains the first turning point, while
the second maximum of the wave function only starts to fit onto the lattice
for $L>1.6$~fm,
giving rise to the second turning point.

\begin{figure}[htp]
\epsfxsize=14truecm
\epsffile{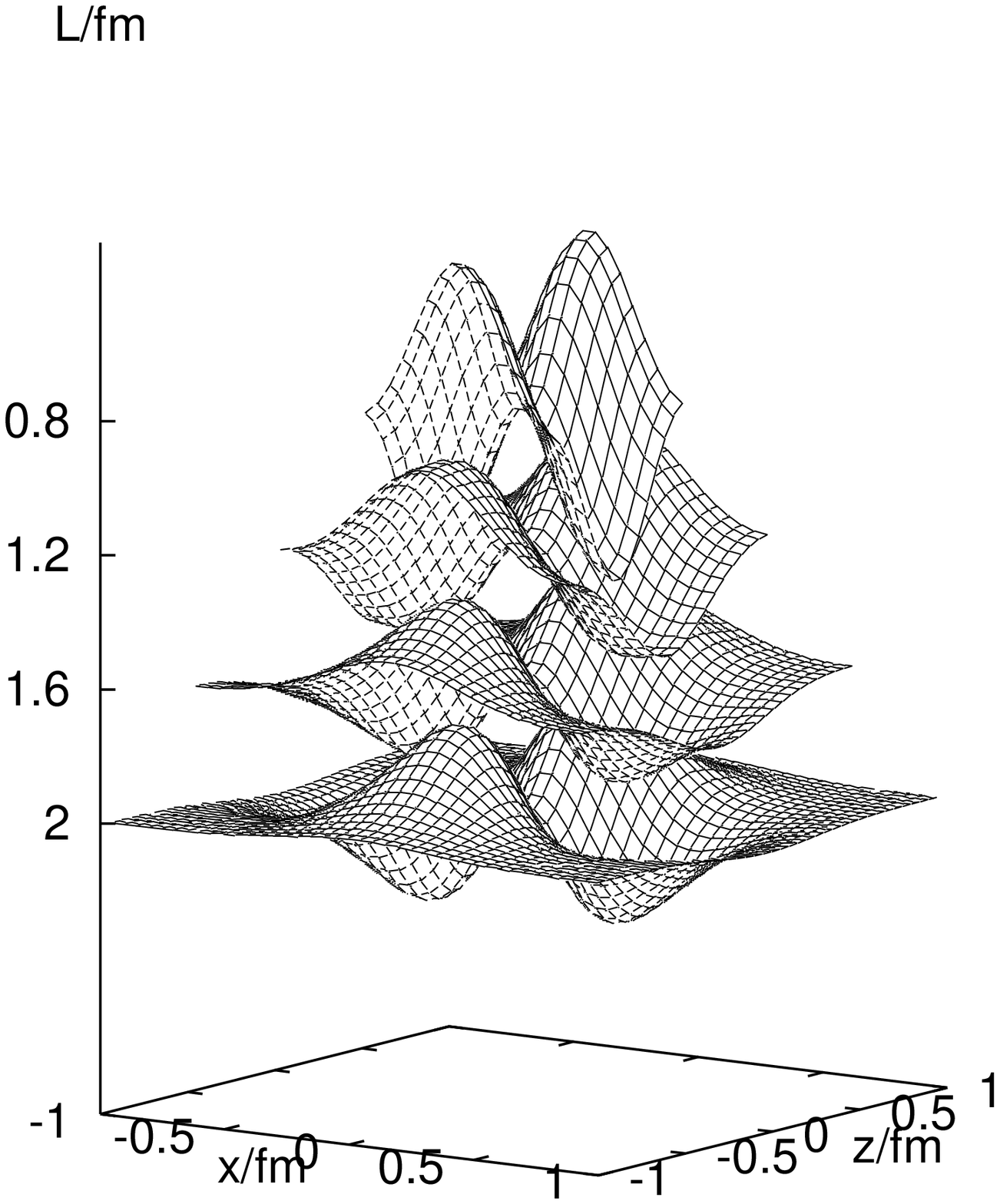}
\caption{A $1D$ wave function in the $E$ representation
at $y=0$ for various
lattice volumes.}
 \label{fse4}
\end{figure}

\begin{figure}[htp]
\epsfxsize=14truecm
\epsffile{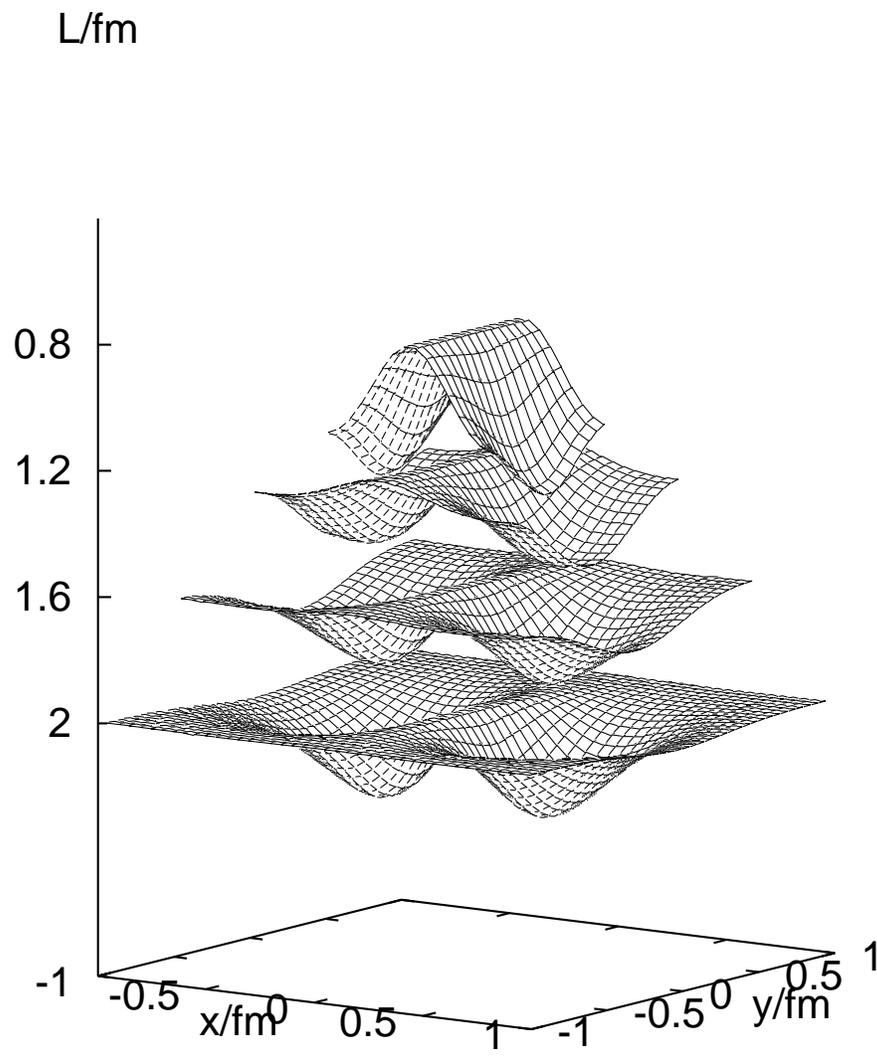}
\caption{Same as Fig.~\protect\ref{fse4} at $z=0$.}
 \label{fse5}
\end{figure}

\begin{figure}[htp]
\epsfxsize=14truecm
\epsffile{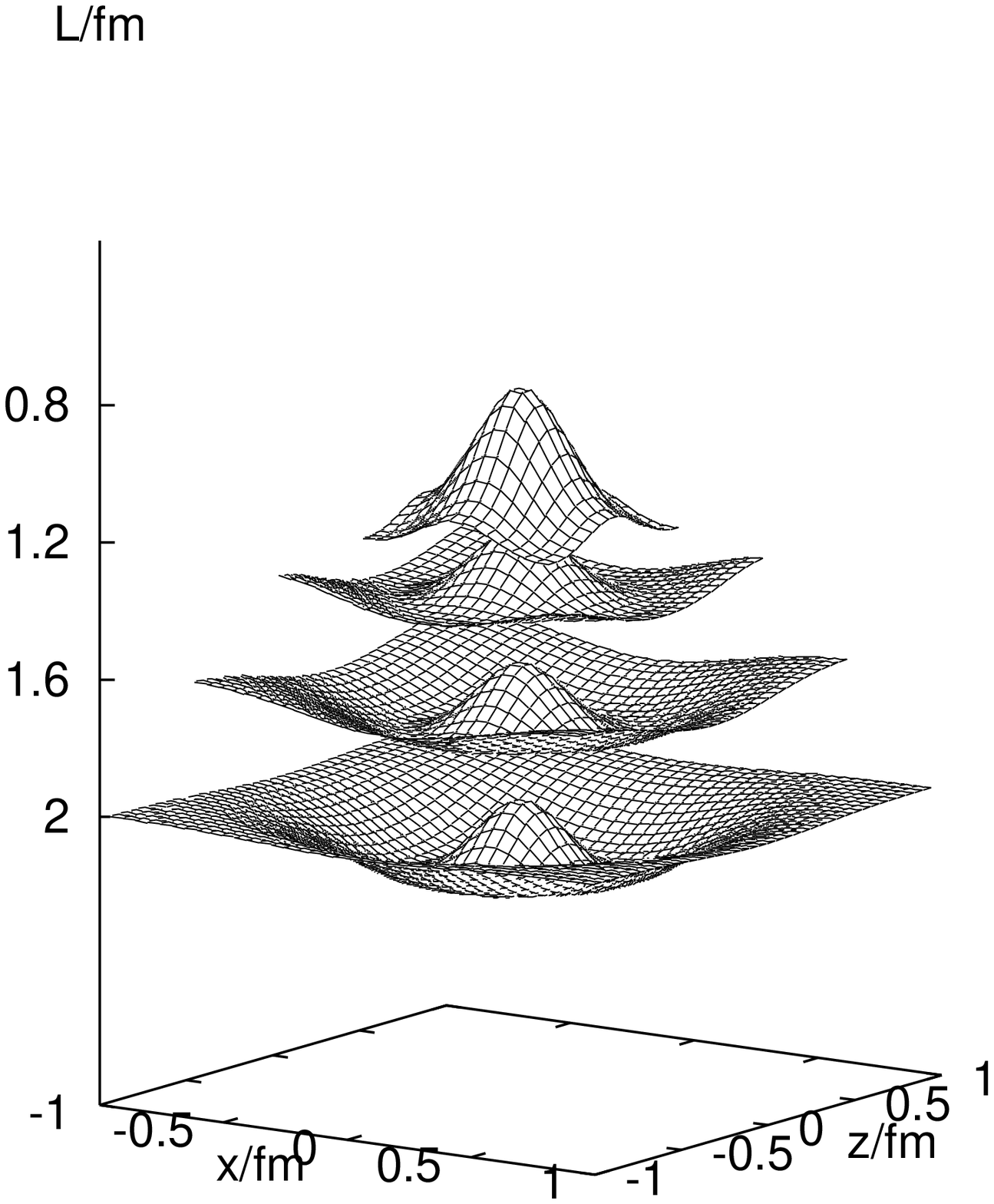}
\caption{Another $1D$ wave function in the $E$ representation
at $y=0$ for various
lattice volumes.}
 \label{fse6}
\end{figure}

\begin{figure}[htp]
\epsfxsize=14truecm
\epsffile{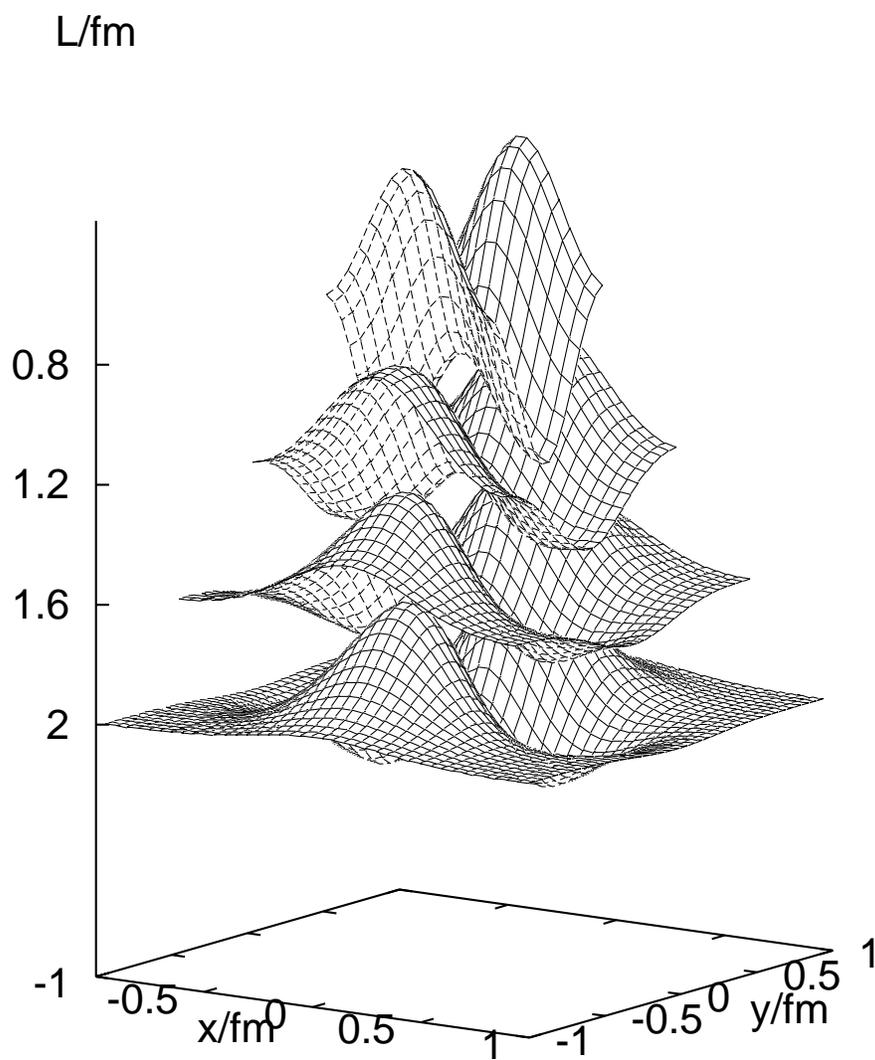}
\caption{Same as Fig.~\protect\ref{fse6} for $z=0$.}
 \label{fse7}
\end{figure}

\begin{figure}[htp]
\epsfxsize=14truecm
\epsffile{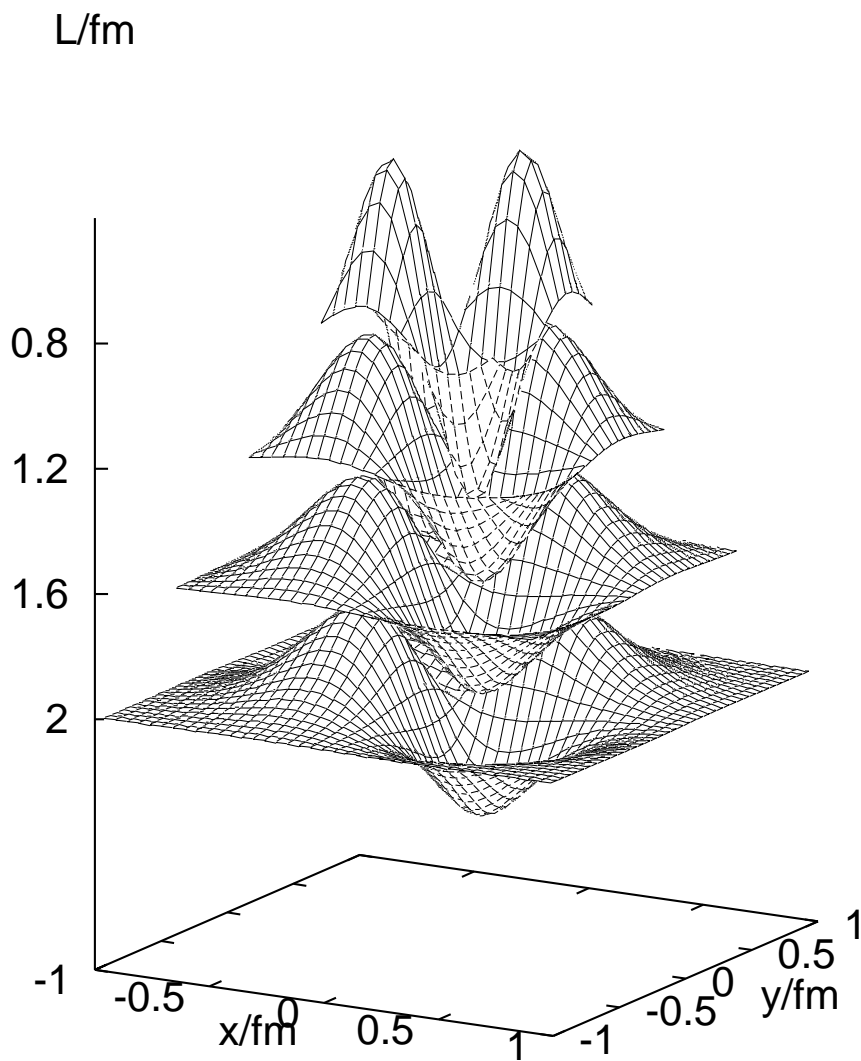}
\caption{A $1D$ wave function in the $T_2$ representation
at $z=0$ for various
lattice volumes. This wave function vanishes
at $x=0$ and $y=0$, i.e.\ within the
$x-z$ and $y-z$ planes.}
\label{fse8}
\end{figure}

In Figs.~\ref{fse4} and \ref{fse5} we display cross sections through
the $1D$ wave function in the $E$ representation within the
$x-z$ and $x-y$ planes, respectively. 
In Figs.~\ref{fse6} and \ref{fse7}, we show another $1D$ wave
function in the $E$
representation.
The energies of these two states are identically degenerate
on all volumes. Finally, in Fig.~\ref{fse8}, we display 
a $1D$ wave function in the $T_2$ representation. The wave function
vanishes at $x=0$ and $y=0$. Interestingly, Fig.~\ref{fse8}
looks very much like a
rotated version of Fig.~\ref{fse4}.
On any finite lattice, we find the $1T_2$ energy
level to differ from that of $1E$, even after the continuum extrapolation.

\subsection{Finite lattice spacing effects}
We investigate the restoration of the continuum symmetry as
the lattice spacing is reduced. We pay particular interest to
states within different representations of the cubic symmetry group $O_h$
that are expected to contain the same $O(3)$ spin content. For instance, at
a lattice spacing $a=0.1$~fm and infinite volume we
find the $1T_2$ state to be by 4~MeV heavier than the $1E$ state.
Unlike FSE, finite $a$ effects
critically depend on the form of the action employed. With our
discretisation of the Laplacian, we would expect order $a^2$ effects
to leading order. Note that in our simulation, we have used a
continuum parametrisation of the static QCD potential. In a lattice
simulation, we would only obtain such a potential with a perfect
gauge action. In general, one would expect the static potential to
suffer under discretisation errors too. Insofar, the
results stated here should be understood as mere lower limits on
the discretisation errors of ``real'' lattice simulations
of the $\Upsilon$ spectrum.

\begin{figure}[htp]
\epsfxsize=14truecm
\epsffile{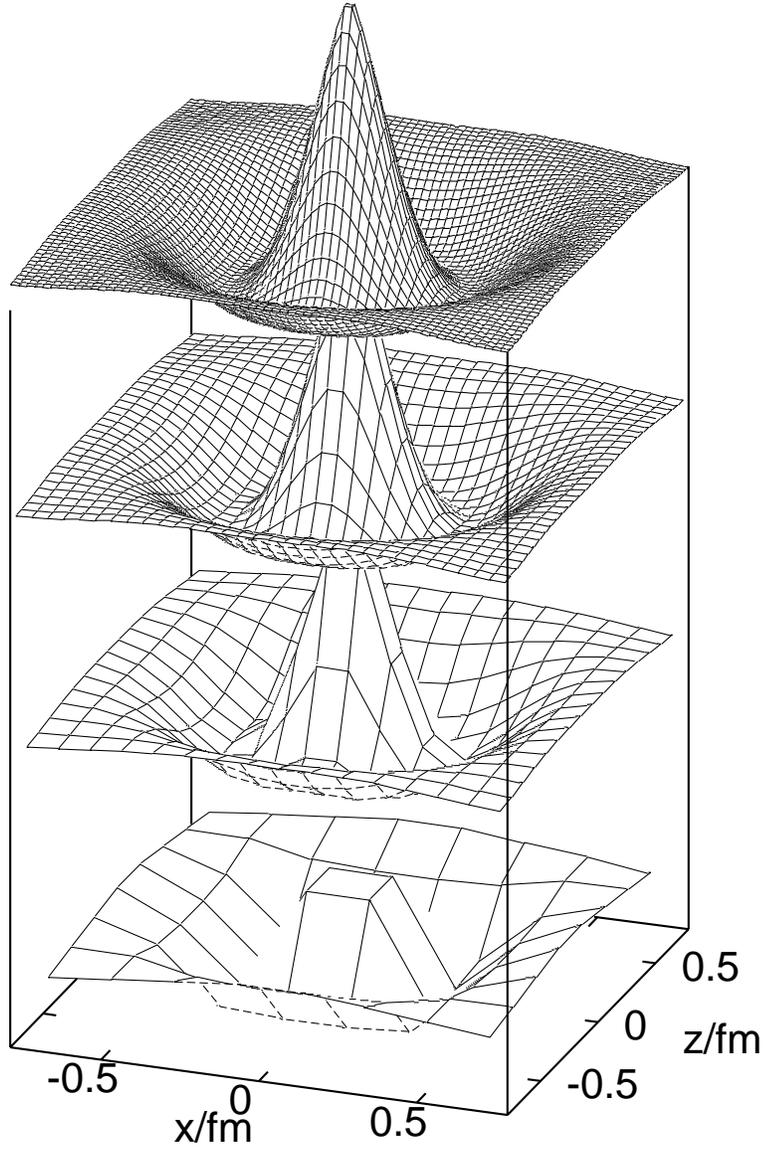}
\caption{The $3S$ wave function on lattices of extent $L=2$~fm and
lattice spacings $a=0.025,0.05,0.1$ and 0.2~fm, respectively.}
\label{finite3s}
\end{figure}

\begin{figure}[htp]
\epsfxsize=11truecm
\centerline{\epsffile{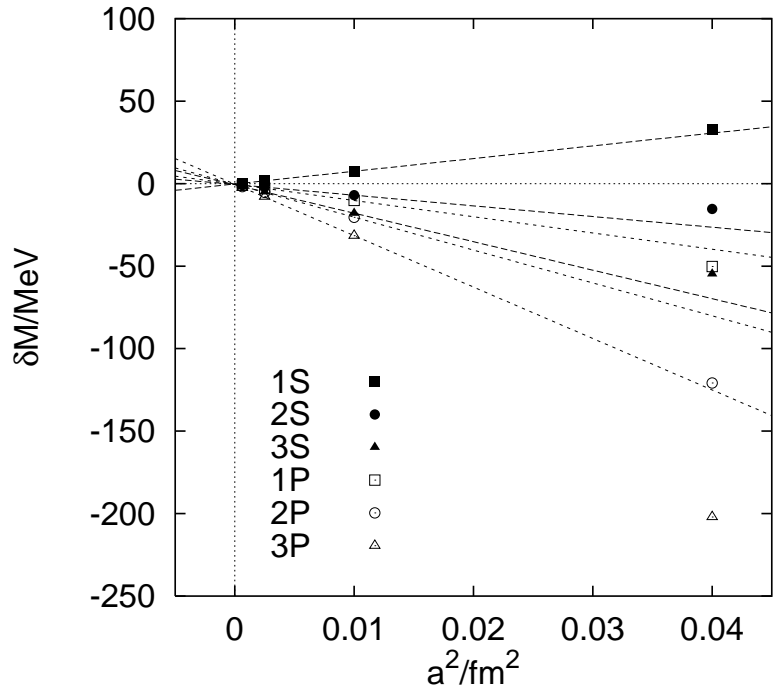}}
\caption{$S$ and $P$ states on lattices of extent 3.6~fm extrapolated
to the continuum limit.}
\label{finitea1}
\end{figure}

\begin{figure}[htp]
\epsfxsize=11truecm
\centerline{\epsffile{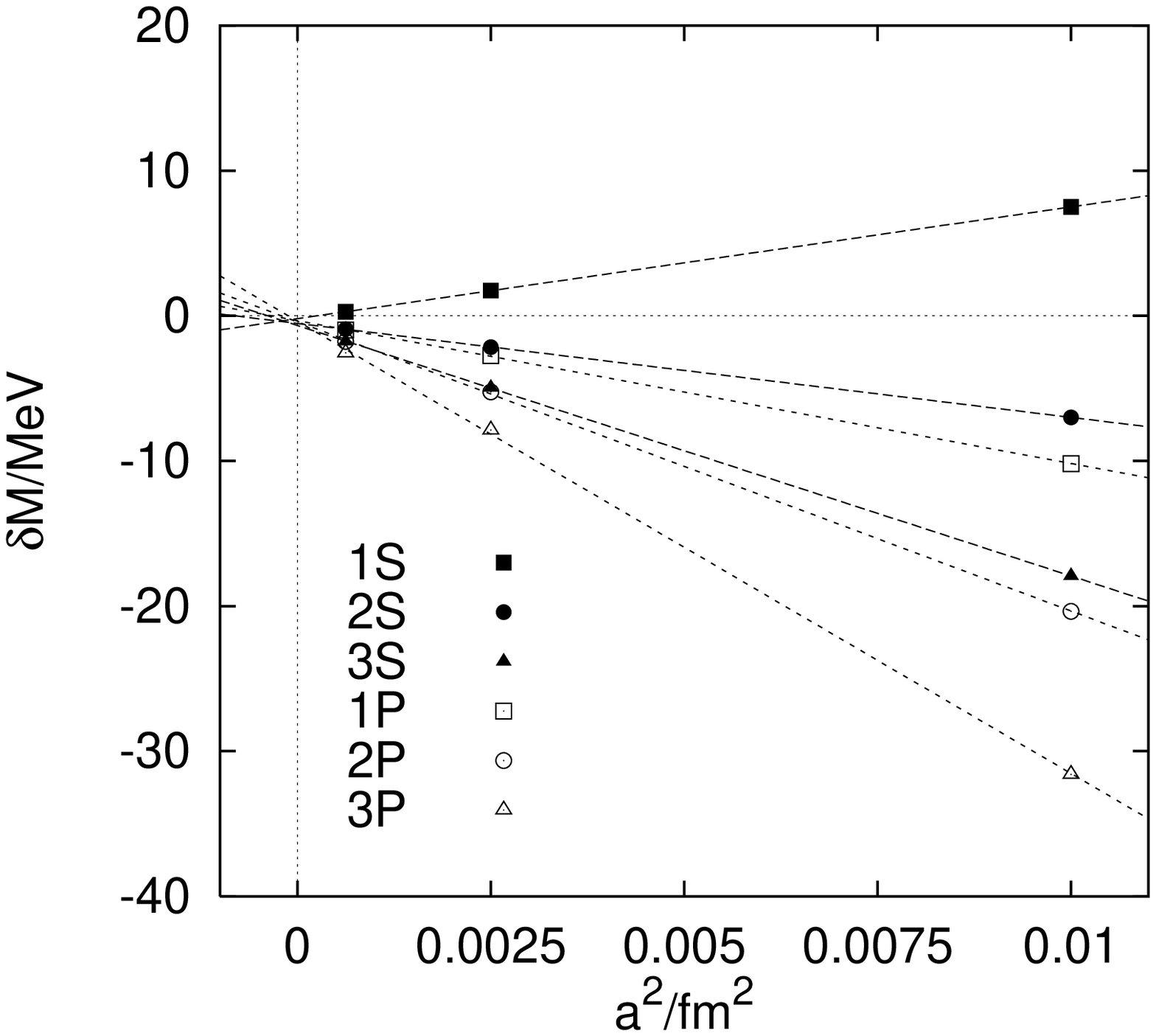}}
\caption{Same as Fig.~\protect\ref{finitea1}.}
\label{finitea2}
\end{figure}

As an illustrative example, in Fig.~\ref{finite3s}, we display the spatial
structure of the $3S$ state for various lattice spacings. While the
general structure of the wave function is still quite well reproduced,
even at $a=0.2$~fm, the peaks themselves are not very well sampled anymore.
In Fig.~\ref{finitea1} we extrapolate $S$ and $P$ states to the continuum
limit. We observe that at finite lattice spacings all levels are
somewhat underestimated --- except for the $1S$ level which turns
out to be
slightly overestimated. Moreover, with increasing radial excitation
$n$, the slope
of the extrapolation in $a^2$ decreases. The lines have been obtained from
fits to the $a=0.025$, 0.05 and 0.1~fm data points. This region is
enlarged in Fig.~\ref{finitea2}.

We interpret the overestimation
of the $1S$ level to be connected to the fact that we placed the origin
--- and therefore the singularity of the Coulomb potential --- at
the centre of an elementary lattice cube. As we increase the lattice spacing,
the $S$ waves sample less and less of the potential around the singular
region and, therefore, we underestimate this negative contribution to the
total energy. The same argument holds had we taken a potential measured
on the lattice, instead of the continuum parametrisation, since in this case
the singularity would have been naturally
cut off by the lattice spacing too. The
decrease of the slope of the extrapolation with increasing number of nodes
of the wave function can be explained by an underestimation of the kinetic
energy due to the reduced curvature on a coarse lattice.

\begin{figure}[htp]
\epsfxsize=11truecm
\centerline{\epsffile{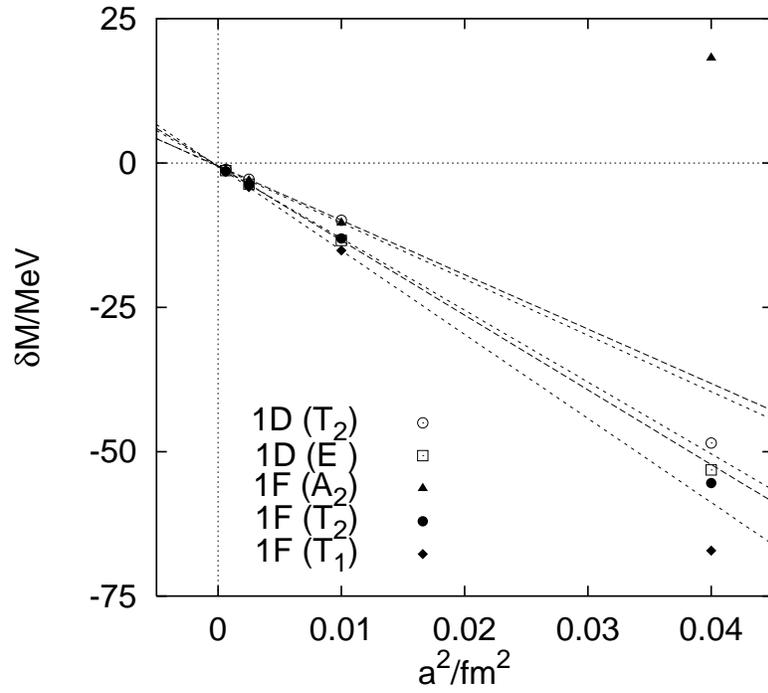}}
\caption{$1D$ and $1F$ states within various lattice representations
extrapolated to the continuum limit on a 3.6 fm lattice.}
\label{finitea3}
\end{figure}

In Fig.~\ref{finitea3}, we investigate how degeneracies of
different representations of $O_h$ that correspond to the same
$O(3)$ spin content become restored in the continuum limit
for $1D$ and $1F$ states. 
The $D$ state in the $T_2$ representation suffers less under both, finite
$a$ effects and FSE than that in the $E$ representation. This is possibly
due to the less complex structure of the wave function (Fig.~\ref{fse8}).
We observe the same for $F$ wave functions in the $A_2$ representation.
While all states are in nice agreement with the leading order $a^2$
expectation up to a resolution $a=0.1$~fm, higher order terms have a
considerable effect at $a=0.2$~fm. This is in particular so for the
$1A_2$ state.

\section{Conclusions}
\label{Summary}
We confirm that ratios of spin averaged bottomonium level splittings
react in a sensitive way towards quenching.
The main origin of this sensitivity is a renormalised
overall-value of the intermediate energy
effective QCD coupling, rather than an
altered running of the coupling between different short distance
scales.
We find that the extent of agreement of spin averaged 
$2S\,1S$ and $1P\,1S$ splittings between the
$\Upsilon$ and $J/\psi$ systems is somewhat accidental. Around the
mass of the charm quark the slope of these splittings as a function
of the inverse quark mass is quite significant but compensated for by
a divergent term that gains influence within the region of the
bottom quark mass.

We have estimated systematic uncertainties of calculations
of quarkonia level splittings based on order $v^4$ NRQCD 
due to ${\mathcal O}(\alpha_s v^4,v^6)$ correction
terms. We end up with estimated accuracies of 1--2~\% and 10~\% on
spin averaged $\Upsilon$ and $J/\psi$ level splittings, respectively.
For the fine structure, our estimates are 35~\% and a factor {\em two},
respectively. The main sources of uncertainty in all cases are radiative
corrections to the matching coefficients of dimension five and six
operators within the effective Lagrangian, rather than
higher order relativistic
effects. Unless non-perturbative matching methods become available,
we have little control over the bottomonium fine structure.
By incorporating existing one- and two-loop
results the uncertainty can be somewhat
reduced. However, a comparison of the predicted fine structure
with experiment indicates that even higher order contributions are
significant and likely to exceed our estimated effect of
25~\%. A slow
convergence of perturbation theory
is indicated by a recent two-loop calculation of $c_F(m)$ too~\cite{amoros}.

We investigated finite volume effects, which we expect to be smaller
for quarkonia than for light mesons. Nonetheless, even for
bottomonia, a linear lattice extent bigger than 2~fm is advisable, at least
if one is interested in radial excitations, such as the $2P$ or $3S$
levels.
Except for the $1S$ state, we failed to find simple parametrisations
of the finite size behaviour of energy levels, based on superpositions
of a power series in the inverse spatial volume, and contributions that are
exponentially suppressed. A general result, however, is
that the coefficients of terms proportional to $L^{-3}$ come out
to be smaller than those that accompany
higher order corrections, which results
in a rapid reduction of FSE, once a critical volume has been by-passed.
We regard this result as relevant for all mesonic physics on the lattice.

Finally, we investigated finite lattice spacing effects. A non-vanishing
spacing $a$ is required in lattice simulations of
effective field theories to provide an ultra violet
cut-off for gluon momenta, which means that the continuum limit cannot be
taken. An unwanted by-product of the lattice regularisation is
the violation of the continuum rotational symmetry and dispersion
relation. By parameterising the potential, obtained at finite lattice spacing,
with a continuous rotationally symmetric curve and subsequently solving
the Hamiltonian in the continuum as well as on discrete lattices, we have
qualitatively investigated the finite $a$ behaviour.
We find it worthwhile to simulate Lattice NRQCD --- where one is confined
to lattice spacings in the region of the inverse quark mass ---
with different gluonic
and fermionic actions and in particular
to realise $D$ waves in both possible
representations of the cubic symmetry group to gain control over
discretisation errors.

Solving the Schr\"odinger-Pauli equation
with lattice potentials on finite volumes provides us with
a powerful tool to predict finite size effects
in lattice simulations and to optimise smearing functions.
The next step will be to extend the method to heavy-light systems
within a Bethe-Salpeter framework which is theoretically not
as rigorously founded as the potential approach for quarkonia
but should
still yield very valuable information on finite size
effects and wave functions, which can then be used in
lattice spectroscopy or determination of decay matrix elements.
Another extension is the inclusion of
retardation effects and incorporation of a running coupling into
the parametrisation.
Work along these lines is in progress.

\acknowledgements
GSB has been funded by the Deutsche Forschungsgemeinschaft
(grants Ba1564/3-1 and Ba1564/3-2) and thanks
the Department of Physics and Astronomy at the University of Glasgow
for providing an inspiring atmosphere when this project was started.
GSB acknowledges helpful discussions with Nora Brambilla, Joan Soto and
Antonio Vairo. PB has been funded by PPARC (grant PP/CBA/62), and wishes
to thank the hospitality of the University of California, Santa Barbara,
where some of this work was carried out.
All computations have been performed on a 64~MB, 133~MHz Pentium PC.
We would like to express our gratitude to everyone who has contributed
to the development of Linux, the GNU compilers and gnuplot.

\begin{table}[h]
\caption{Spin content of states.}
\label{tab:rep}
\begin{center}
\begin{tabular}{ccc}
        boundary ($xz$)    &$O_h$& $l$\\\hline
        $ee$       & $A_1$, $E$  &$0,2,4\ldots$\\
        $eo$       & $T_1$       &$1,3,\ldots$\\
        $oe$       & $T_2$       &$2,3,\ldots$\\
        $oo$       & $A_2$       &$3,6,\ldots$\\
\end{tabular}
\end{center}
\end{table}

\begin{table}[h]
\caption{Smallest lattice extents at which FSE are found to be smaller than
3~MeV.}
\label{tab:fse}
\begin{center}
\begin{tabular}{lcc}
        state&$L_{\mbox{\scriptsize 3~MeV}}/$fm&
$\langle r^2\rangle^{1/2}/$fm\\\hline
        $1S$&1.25&0.25\\
        $1P$&1.55&0.41\\
        $2S$&1.85&0.52\\
        $1D (T_2)$&1.75&0.54\\
        $1D (E)$&2.0&0.54\\
        $2P$&2.1&0.65\\
        $3S$&2.4&0.74\\
        $3P$&2.55&0.85\\
\end{tabular}
\end{center}
\end{table}



\end{document}